\documentclass[reprint,showpacs,amsmath,amssymb,superscriptaddress]{revtex4}
\usepackage{multirow}
\usepackage{bigstrut}
\usepackage{amssymb}
\usepackage{amsmath}
\usepackage{graphicx}
\usepackage{dcolumn}
\usepackage{bm}
\usepackage{CJK}
\usepackage{relsize}
\usepackage{cancel}
\usepackage{color}

\begin{document}

\title{Theoretical Details of Tunnel Magnetoresistance via inelastic hopping at regime $g\mu B\ll k_B T \ll eV$}

\author{Yang Song}\email{yangsong@pas.rochester.edu}\affiliation{Department of Electrical and Computer Engineering, University of Rochester, Rochester, New York, 14627}

\begin{abstract}
Detailed theoretical derivation is given for the tunnel magnetoresistance via phonon-assisted hopping through an impurity chain under small magnetic field and a large bias window. This derivation provides a rigorous basis for the physical picture of Pauli blockade switch proposed in our previous paper (arXiv:1404.0633). This picture captures the competition of external magnetic field and internal spin interactions in the tunnel barrier, and relies critically on the strong on-site Coulomb correlation at the impurities. The master equations are obtained by deriving the equations of motion for the Green functions at the impurity sites in a slave-boson representation, and utilizing the so-called Langreth theorem to finally express the spin-dependent density matrix in terms of the equilibrium distributions of the contact electrons and of the phonon reservoir.
\end{abstract}
\maketitle

\section{introduction}

Recent experimental evidences of tunnel magnetoresistance (TMR) via phonon-assisted hopping through  impurity chains have suggested a new type of TMR mechanism without the need of a magnetic contact in the tunnel junction \cite{Txoperena_PRL14}. This physics greatly generalizes the TMR due to resonant tunneling in  a ferromagnetic-insulator-nonmagnetic (FIN) junction, which was proposed to explain the puzzling enhanced signals in the three-terminal electrical Hanle effect measurements in the past few years (\cite{Song_PRL14} and references therein). Both TMRs were observed with a relatively weak magnetic field ($\sim$ 1 KG) and large bias window, and at a temperature that is not too low, i.e., often at  regime $g\mu B\ll k_B T \ll eV$. The relative TMR is often around $0.1\sim 1\%$, although in the special case of one-dimensional atomic or molecular chain setting the TMR could take over as the total tunnel resistance \cite{Mahato_Science13}. We have envisioned a novel type of nanometer sized 1D memory cell based on this new TMR effect \cite{Song_PRL14}.

Both of the above mentioned TMR mechanisms rely essentially on the strong on-site Coulomb correlation $U$ at the impurity site to effectively break the spin degeneracy, i.e, to make the spin splitting scale with $U\gg eV$ rather than with the small $g\mu B$. Both mechanisms operate under a competition of the externally applied magnetic field and an effective internal field on the impurities due to the spin interactions such as hyperfine or exchange. The Pauli blockade between the magnetic contact and the impurity or between impurities is opened or reinforced in response to the total fields. For the tunnel junction with a magnetic contact, the spin-polarized tunneling leads to an additional field anisotropy of the TMR effect. As a result, apart from the spin-dependent contact in the FIN case, both TMRs rely on similar barrier impurities that with one of the occupiable energy levels inside the bias window \cite{Song_PRL14, Txoperena_PRL14}. The complexity, however, dramatically increases in the derivation of the TMR with phonon-assisted hopping, considering the number of occupation states of the impurities chains and the involvement of the phonon reservoir. Moreover, special attention is needed  for the coherent off-diagonal terms during the phonon-assisted hopping between impurities. These often-ignored coherent terms are imposed and protected by the symmetry in the spin subspace of the system Hamiltonian, and may lead to observable effect.

It is the purpose of the this manuscript to present the detailed theoretical derivation for the general and more complicated case of hopping-assisted TMR. Not only this derivation puts the physical picture proposed in Ref.~\cite{Txoperena_PRL14} on a solid basis, subtle effects like the coherent terms in the phonon-assisted hopping mentioned above appear thanks to this rigorous derivation. With the increase of the temperature, bias  or the tunnel barrier thickness, the hopping-assisted tunneling is expected to dominate the direct or resonant tunneling until the Mott variable range hopping kicks in for much thicker barriers \cite{Glazman_phonon_JETP88, Xu_PRB95}. Therefore this derivation should be very relevant in a large number of tunneling situations. It will be easily realized that in this TMR mechanism the key process is a hopping through a A-B sequence, which is the process we focus on below. [A(B) type is defined as the first (second) electron filing level of the impurity is within the bias window.]  The TMR effect remains for a hopping chain with more than 2 impurities, as long as there is an A-B sequence within it.

The rest of the manuscript is organized as follows. Sec.~\ref{sec:Hamiltonian} introduces the Hamiltonian of the system. Sec.~\ref{sec:slave-boson} provides various  operators of the impurity density matrix within the restricted occupation space in a slave-boson representation. Sec.~\ref{sec:derive-master} is the key derivation section and derives the master equations all the way from the basic equations of motion, utilizing the so-called Langreth theorem. Sec.~\ref{sec:A-B chain} shows the master equations and their analytical solutions for the important A-B chain which is responsible for the TMR effect. The symmetry of spin subspace and the resulting coherent terms are discussed. Sec.~\ref{sec:other chains} repeats the procedure for other two-impurity chains, including B-A, A-A and B-B chains. Finally, Sec.~\ref{sec:summary} summarizes it briefly.

\section{Hamiltonian of the system}\label{sec:Hamiltonian}

The system Hamiltonian of a tunneling junction with two impurities and electron-phonon (e-ph) interaction is
\begin{eqnarray}
H &=& \sum_{\ell\mathbf{k}\sigma} \varepsilon_{\ell\mathbf{k}\sigma} n_{\ell\mathbf{k}\sigma}
+ \sum_{\ell \sigma} \left[(E_{d\ell} + \sigma E_{B\ell} \cos\theta_\ell) n_{d\ell \sigma} +
E_{B\ell} \sin\theta_\ell d^\dag_{\ell\sigma} d_{\ell\bar\sigma}\right] +\sum_\ell U_\alpha n_{d\ell \uparrow}  n_{d\ell \downarrow}
\nonumber\\
&&
+ \sum_{\ell\mathbf{k}\ell'\sigma} \left[V_{\ell\mathbf{k}\ell'\sigma} \mathbf{k}^\dag_{\ell\sigma} d_{\ell'\sigma}  +\textrm{H.c.} \right]
+ \sum_{\sigma}  ( V_{dd} d^\dag_{L\sigma}d_{R\sigma} +\textrm{H.c.} )
\nonumber\\
&&+ \sum_{\mathbf{q}}\varepsilon_{\mathbf{q}} \mathbf{q}^\dag \mathbf{q}
+ \sum_{\ell \mathbf{q}\sigma} d^\dag_{\ell\sigma} d_{\ell\sigma} (\alpha_{\ell \mathbf{q}} \mathbf{q}^\dag +\alpha^*_{\ell \mathbf{q}} \mathbf{q}).
\end{eqnarray}
where without loss of generality, we are allowed to set $V_{\ell\mathbf{k}\ell'\sigma}$  real. $\mathbf{k}$, $d$ and $\mathbf{q}$ denote the contact electron, impurity electron and phonon  respectively.  $\ell=L,R$  or $+(-)1$ denote the left and right sides.  $\sigma=\uparrow,\downarrow$ or $+(-)1$ denote the spin. $\bar{\sigma}, \bar{\alpha}$ are the opposite of $\sigma, \alpha$. $E_{B\ell}=g\mu B_\ell/2$ is the magnetic ($B$) field at the $\ell$th impurity site, with orientation $\theta_\ell$ and $\phi_\ell=0$ ($xz$ plane is spanned by the two $\mathbf{B}$ directions).    For simplicity,  we do not consider the Coulomb repulsion between two impurities, that is, the energy levels of each impurity is not related to how the other impurity is filled (considering this inter-impurity repulsion will produces more types of two-impurity chain than A-A, B-B, A-B, B-A; see Ref.~\cite{Bahlouli_PRB94} for specifics. It does not change the results qualitatively for our purpose.). $\alpha_{\ell\mathbf{q}} = i \Xi_\ell q/\sqrt{2\rho \varepsilon_\mathbf{q}} \exp(i\mathbf{q}\mathbf{R}_\ell)$, where $\Xi_\ell$ and $\mathbf{R}_\ell$ denote deformation potential and position of $\ell$th impurity, and $\rho$ is the insulator mass density. The phonon-assisted hopping is assumed to conserve spin and the weak phonon induced spin flip is neglected.

In the following, we deal with two normal metal contacts ($\varepsilon_{\ell\mathbf{k}\sigma}=\varepsilon_{\ell\mathbf{k}}$) and two-impurity bridge situation, so that $V_{\ell\mathbf{k}\ell'\sigma}= \delta_{\ell,\ell'} V_{\ell\mathbf{k}}$. For electron-phonon interaction, we do the usual unitary transformation and follow Ref.~\cite{Glazman_phonon_JETP88} to keep only linear terms in $\alpha_{\ell \mathbf{q}}$ (weak e-ph interaction). We are presently interested in the situation where the phonon-assisted hopping dominates the resonant tunneling. Then the Hamiltonian becomes
\begin{eqnarray}\label{eq:Hamiltonian_approx}
H &=& \sum_{\ell\mathbf{k}\sigma} \varepsilon_{\ell\mathbf{k}} n_{\ell\mathbf{k}\sigma}
+ \sum_{\ell \sigma} \left[(E_{d\ell} + \sigma E_{B\ell} \cos\theta_\ell) n_{d\ell \sigma} +
E_{B\ell} \sin\theta_\ell d^\dag_{\ell\sigma} d_{\ell\bar\sigma}\right] +\sum_\ell U_\alpha n_{d\ell \uparrow}  n_{d\ell \downarrow}
\nonumber\\
&&+ \sum_{\ell\mathbf{k}\sigma} \left(V_{\ell\mathbf{k}} \mathbf{k}^\dag_{\ell\sigma} d_{\ell\sigma}  +\textrm{H.c.} \right)
+ iV_{dd} \sum_{\sigma} (  d^\dag_{L\sigma}d_{R\sigma} +d^\dag_{R\sigma}d_{L\sigma} ) \sum_{\mathbf{q}} (\lambda_\mathbf{q} \mathbf{q}^\dag +\lambda^*_\mathbf{q} \mathbf{q})
+ \sum_{\mathbf{q}}\varepsilon_{\mathbf{q}} \mathbf{q}^\dag \mathbf{q}
,
\end{eqnarray}
where without loss of generality, we set $V_{dd}$ pure imaginary, and $\lambda_\mathbf{q}=i(\alpha_{R\mathbf{q}}-\alpha_{L\mathbf{q}} )/\varepsilon_\mathbf{q}$.

\section{slave-boson operators}\label{sec:slave-boson}

The auxiliary particle approach was pioneered by Abrikosov (pseudofermions) \cite{Abrikosov_Phy65}, and later by Barnes \cite{Barnes_JPhys76} and Coleman \cite{Coleman_PRB84}. By using slave-particle representation, the occupation constraint under on-site Coulomb interaction become precise and convenient [for example see Eq.~(\ref{eq:constraint}) in the following]. The slave-boson approach was first used in Ref.~\cite{Zou_PRB88} for finite-$U$ Hubbard model. Here we generalize the definition and (anti)commutation rules used in Ref.~\cite{LeGuillou_PRB95}, for our case of a two-impurity chain.

To satisfy the correct quantization, i.e. (anti)commutation rules, the real electron annihilation operators can be expressed in terms of pseudo-boson and pseudo-fermion states (as $|n_{L}  n_{R}\rangle$ where $n=\{0,\uparrow,\downarrow, 2\}$ are four possible states at the impurity) as follows,
\begin{subequations} \label{eq:slave_boson_full}
\begin{eqnarray}
d_{L\sigma}&=&|00\rangle \langle \sigma 0|+ \sigma|\bar \sigma0\rangle \langle 2 0| + \tau_L  \sum_{\sigma'}\left(|0 \sigma'\rangle \langle\sigma\sigma'| + \sigma|\bar\sigma \sigma'\rangle \langle 2\sigma'|\right) + |02\rangle \langle\sigma2| + \sigma|\bar\sigma 2\rangle \langle 22|,
\\
d_{R\sigma}&=&|00\rangle \langle  0\sigma|+ \sigma|0\bar \sigma\rangle \langle 02| + \tau_R  \sum_{\sigma'}\left(|\sigma'0\rangle \langle\sigma'\sigma| + \sigma|\sigma' \bar\sigma\rangle \langle \sigma'2|\right) + |20\rangle \langle2\sigma| + \sigma| 2\bar\sigma\rangle \langle 22|,
\end{eqnarray}
\end{subequations}
together with the quantization rules that $\langle n_{L}  n_{R}|m_{L}  m_{R}\rangle=\delta_{n_L m_L}\delta_{n_R m_R}$. Note that in the large $U$ limit, Eqs.~(\ref{eq:slave_boson_full}) can be approximated as 
\begin{subequations} \label{eq:slave_boson_largeU}
\begin{eqnarray}
d_{L\sigma}=\tau_L  \sum_{\sigma'}|0 \sigma'\rangle \langle\sigma\sigma'| + |02\rangle \langle\sigma2|, 
\quad
d_{R\sigma}= \sigma|0\bar \sigma\rangle \langle 02| + \tau_R  \sum_{\sigma'} \sigma|\sigma' \bar\sigma\rangle \langle \sigma'2| ,
\end{eqnarray}
and we are led to the constraint
\begin{eqnarray}\label{eq:constraint}
1=\rho_{02}+\sum_\sigma (\rho_{0\sigma}+ \rho_{\sigma\bar\sigma} +\rho_{\sigma2} +\rho_{\sigma\sigma}).
\end{eqnarray}
\end{subequations}

Note that the coefficients $\sigma$, $\tau_L=-\tau_R$ is necessary for all the (anti)commutation rules to be satisfied. We take $\tau_L=1$ in the following. Now the terms in the Hamiltonian [Eq.~(\ref{eq:Hamiltonian_approx})] can be readily constructed in the new representation.
\begin{subequations}
\begin{eqnarray}
n_{L\sigma}&=&|\sigma0\rangle \langle \sigma 0|+ |20\rangle \langle 2 0| +  \sum_{\sigma'}\left(|\sigma \sigma'\rangle \langle\sigma\sigma'| + \sigma|2 \sigma'\rangle \langle 2\sigma'|\right) + |\sigma2\rangle \langle\sigma2| + |2 2\rangle \langle 22|
\\
n_{R\sigma}&=&|0\sigma\rangle \langle  0\sigma|+ |02\rangle \langle 02| +  \sum_{\sigma'}\left(|\sigma'\sigma\rangle \langle\sigma'\sigma| + \sigma|\sigma' 2\rangle \langle \sigma'2|\right) + |2\sigma\rangle \langle2\sigma| + | 22\rangle \langle 22|,
\\
d^\dag_{L\sigma}d_{R\sigma}&=&|\sigma0\rangle \langle  0\sigma|+ \tau_R\sigma |20\rangle \langle\bar\sigma\sigma|+ \tau_L\sigma |\sigma\bar\sigma\rangle \langle02| - | 2\bar\sigma\rangle \langle \bar\sigma2|,
\\
d^\dag_{R\sigma}d_{L\sigma}&=&|0\sigma\rangle \langle \sigma 0|+ \tau_L\sigma |02\rangle \langle\sigma\bar\sigma|+ \tau_R\sigma |\bar\sigma\sigma\rangle \langle20| - | \bar\sigma2\rangle \langle 2\bar\sigma|,
\\
d^\dag_{L\sigma}d_{L\bar\sigma}&=&|\sigma0\rangle \langle \bar\sigma 0|+ \sum_{\sigma'} |\sigma\sigma'\rangle \langle\bar\sigma\sigma'|+  |\sigma2\rangle \langle\bar\sigma2|,
\\
d^\dag_{R\sigma}d_{R\bar\sigma}&=&|0\sigma\rangle \langle 0\bar\sigma |+ \sum_{\sigma'} |\sigma'\sigma\rangle \langle\sigma'\bar\sigma|+  |2\sigma\rangle \langle2\bar\sigma|.
\end{eqnarray}
\end{subequations}

\section{From equations of motion to master equations using Langreth theorem}\label{sec:derive-master}

In order to write down the master equation for the density matrix elements in terms of themselves and equilibrium distribution of the contact electrons and of the phonon reservoir, we start with the basic equation of motion for operators in the Heisenberg picture. In the following, let us focus on the tricky terms that are induced by the the electron-phonon interaction [$iV_{dd} \sum_{\sigma} (d^\dag_{L\sigma}d_{R\sigma} +d^\dag_{R\sigma}d_{L\sigma} ) \sum_{\mathbf{q}} (\lambda_\mathbf{q} \mathbf{q}^\dag +\lambda^*_\mathbf{q} \mathbf{q})$ in Eq.~(\ref{eq:Hamiltonian_approx})]. We define density operators $\hat\rho^{n_{L}  n_{R}}_{m_{L}  m_{R}}\equiv |n_{L}  n_{R}\rangle \langle m_{L}  m_{R}|$ and $\hat\rho_{n_{L}  n_{R}}\equiv \hat\rho^{n_{L}  n_{R}}_{n_{L}  n_{R}}$. For operator $\hat\rho_{\uparrow\downarrow}$,
\begin{eqnarray}\label{eq:rate_rho_ud}
\hbar \frac{d}{dt}\hat\rho_{\uparrow\downarrow} = -V_{dd}\sum_\mathbf{q} (\lambda_\mathbf{q} \mathbf{q}^\dag +\lambda^*_\mathbf{q} \mathbf{q})(\hat\rho^{02}_{\uparrow\downarrow}-\hat\rho^{\uparrow\downarrow}_{02})+ ....
\end{eqnarray}
For the ensemble average of the correlation term $\langle \mathbf{q}^\dag(t') \hat\rho^{02}_{\uparrow\downarrow}(t)\rangle$, it has
\begin{eqnarray}
(-i\hbar \frac{d}{dt'}-\varepsilon_\mathbf{q})\langle \mathbf{q}^\dag(t') \hat\rho^{02}_{\uparrow\downarrow}(t)\rangle = iV_{dd} \lambda_\mathbf{q}^* \langle \sum_\sigma (d^\dag_{L\sigma}d_{R\sigma} +d^\dag_{R\sigma}d_{L\sigma} )(t')\hat\rho^{02}_{\uparrow\downarrow}(t)\rangle.
\end{eqnarray}
The first operator on the LHS is the inverse of phonon Green function operator ($g_\mathbf{q}$). By the so-called Langreth theorem \cite{Langreth_chap76,Haug_Jauho_book08}, one has
\begin{eqnarray}
\langle \langle  \hat\rho^{02}_{\uparrow\downarrow},\mathbf{q}^\dag\rangle\rangle^<(\varepsilon) =
iV_{dd} \lambda_\mathbf{q}^* [\langle \langle \hat\rho^{02}_{\uparrow\downarrow}, \sum_\sigma (d^\dag_{L\sigma}d_{R\sigma} +d^\dag_{R\sigma}d_{L\sigma} )\rangle\rangle^R(\varepsilon) g^<_\mathbf{q}(\varepsilon)
 +
\langle \langle \hat\rho^{02}_{\uparrow\downarrow}, \sum_\sigma (d^\dag_{L\sigma}d_{R\sigma} +d^\dag_{R\sigma}d_{L\sigma} )\rangle\rangle^<(\varepsilon) g^A_\mathbf{q}(\varepsilon)].
\end{eqnarray}

Therefore, the ensemble average of two terms on RHS of Eq.~(\ref{eq:rate_rho_ud}), for weak contact-impurity coupling, is,
\begin{eqnarray}\label{eq:q_rho02ud}
-V_{dd}\sum_\mathbf{q} (\lambda_\mathbf{q} \langle \mathbf{q}^\dag \hat\rho^{02}_{\uparrow\downarrow}\rangle- \lambda^*_\mathbf{q} \langle \mathbf{q} \hat\rho_{02}^{\uparrow\downarrow}\rangle)
=
2{\rm Re}\left[-V_{dd}\sum_\mathbf{q} \lambda_\mathbf{q} \langle \mathbf{q}^\dag \hat\rho^{02}_{\uparrow\downarrow}\rangle\right]
= \Gamma_d [ \rho_{02}n_\mathbf{q}-(\rho_{\uparrow\downarrow}-{\rm Re} \rho^{\downarrow\uparrow}_{\uparrow\downarrow})(n_\mathbf{q}+1)].
\end{eqnarray}
where $\Gamma_d = 2\pi \sum_\mathbf{q}|V_{dd}\lambda_q|^2\delta(|E_{dL}-E_{dR}|-\varepsilon_{\mathbf{q}})$ and $\rho^{n_{L}  n_{R}}_{n_{L}  n_{R}}=\langle \hat\rho^{n_{L}  n_{R}}_{n_{L}  n_{R}} \rangle$. Besides the regular identities $G^R-G^A=G^>-G^<$ and $\int d\varepsilon (G^R+G^A)=0$, we have used the rules for lesser and greater Green functions for phonons and boson particles, which are different from that of the usual electron operators,
\begin{subequations}
\begin{eqnarray}
g^{</>}_\mathbf{q}&=&-2\pi i[(n_\mathbf{q}+1)\delta(\varepsilon\pm\varepsilon_\mathbf{q})
+n_\mathbf{q}\delta(\varepsilon\mp\varepsilon_\mathbf{q})],\label{eq:q_lesser}
\\
\langle \langle  \hat\rho^{02}_{\uparrow\downarrow},\mathbf{q}^\dag\rangle\rangle^<
 &= &
 -2\pi i \langle \mathbf{q}^\dag \hat\rho^{02}_{\uparrow\downarrow} \rangle \delta(\varepsilon-\Delta \varepsilon_d),\label{eq:rho02ud_q_lesser}
\\
\langle \langle \hat\rho^{02}_{\uparrow\downarrow}, \sum_\sigma (d^\dag_{L\sigma}d_{R\sigma} +d^\dag_{R\sigma}d_{L\sigma} )\rangle\rangle^<
&=&
 -2\pi i \langle  \sum_\sigma (d^\dag_{L\sigma}d_{R\sigma} +d^\dag_{R\sigma}d_{L\sigma} )\hat\rho^{02}_{\uparrow\downarrow} \rangle \delta(\varepsilon-\Delta \varepsilon_d)
\nonumber \\
 &= &-2\pi i (\rho_{\uparrow\downarrow}-\rho^{\downarrow\uparrow}_{\uparrow\downarrow})\delta(\varepsilon-\Delta \varepsilon_d),\label{eq:rho02ud_dLR_lesser}
 \\
\langle \langle \hat\rho^{02}_{\uparrow\downarrow}, \sum_\sigma (d^\dag_{L\sigma}d_{R\sigma} +d^\dag_{R\sigma}d_{L\sigma} )\rangle\rangle^>
&=&
 -2\pi i \langle  \hat\rho^{02}_{\uparrow\downarrow} \sum_\sigma (d^\dag_{L\sigma}d_{R\sigma} +d^\dag_{R\sigma}d_{L\sigma} )\rangle \delta(\varepsilon-\Delta \varepsilon_d)
\nonumber \\
 &= &-2\pi i \rho_{02}\delta(\varepsilon-\Delta \varepsilon_d),\label{eq:rho02ud_dLR_greater}
\end{eqnarray}
\end{subequations}
where $\Delta \varepsilon_d=\varepsilon_{dL}-\varepsilon_{dR}$. We emphasize two noteworthy points. Firstly the negative sign for lesser Green function in Eqs.~(\ref{eq:q_lesser})-(\ref{eq:rho02ud_dLR_lesser}) is critical for reaching the correct result in Eq.~(\ref{eq:q_rho02ud}) and is opposite to that for fermion operators, which is due to the fact that each operator in the two-time Green functions is either for phonon or for two fermions. Secondly in reaching Eq.~(\ref{eq:rate_rho_ud}) we neglect off-diagonal matrix element between the occupiable states [the 9 states, for example, contained in Eqs.~(\ref{eq:slave_boson_largeU})] and subspace outside it. This is justified by that these off-diagonal elements will be scaled down by a factor of $1/U\ll 1$ where $U$ is our largest energy scale, and simplifies the situation from a even more complicated one.

With the same procedure, one has
\begin{eqnarray}
-V_{dd}\sum_\mathbf{q} (-\lambda_\mathbf{q} \langle \mathbf{q}^\dag \hat\rho_{02}^{\uparrow\downarrow}\rangle+ \lambda^*_\mathbf{q} \langle \mathbf{q} \hat\rho^{02}_{\uparrow\downarrow}\rangle)
= \Gamma_d [ \rho_{02}n_\mathbf{q}-(\rho_{\uparrow\downarrow}-{\rm Re} \rho^{\downarrow\uparrow}_{\uparrow\downarrow})(n_\mathbf{q}+1)].
\end{eqnarray}
So Eq.~(\ref{eq:rate_rho_ud}) leads to
\begin{subequations}
\begin{eqnarray}\label{eq:rate_rho_ud_final}
\hbar \frac{d}{dt}\rho_{\uparrow\downarrow} =
2\Gamma_d [ \rho_{02}n_\mathbf{q}-(\rho_{\uparrow\downarrow}-{\rm Re} \rho^{\downarrow\uparrow}_{\uparrow\downarrow})(n_\mathbf{q}+1)]+ ....
\end{eqnarray}
Similarly one has
\begin{eqnarray}
\hbar \frac{d}{dt}\rho_{\downarrow\uparrow}
&=&
2\Gamma_d [ \rho_{02}n_\mathbf{q}-(\rho_{\downarrow\uparrow}-{\rm Re} \rho^{\downarrow\uparrow}_{\uparrow\downarrow})(n_\mathbf{q}+1)]+ ....
\\
\hbar \frac{d}{dt}\rho^{\downarrow\uparrow}_{\uparrow\downarrow}
&=&
\Gamma_d [ -2\rho_{02}n_\mathbf{q}+(\rho_{\uparrow\downarrow}+\rho_{\downarrow\uparrow}
- 2\rho^{\downarrow\uparrow}_{\uparrow\downarrow})(n_\mathbf{q}+1)]+ ....
\\
\hbar \frac{d}{dt}\rho_{02}
&=&
2\Gamma_d [ -2\rho_{02}n_\mathbf{q}+(\rho_{\uparrow\downarrow}+\rho_{\downarrow\uparrow}
- 2{\rm Re}\rho^{\downarrow\uparrow}_{\uparrow\downarrow})(n_\mathbf{q}+1)]+ ....
\\
\hbar \frac{d}{dt}\rho^{\uparrow\downarrow}_{\uparrow\uparrow}=-\hbar \frac{d}{dt}\rho^{\downarrow\uparrow}_{\uparrow\uparrow}
&=&
\Gamma_d ( -\rho^{\uparrow\downarrow}_{\uparrow\uparrow}+\rho^{\downarrow\uparrow}_{\uparrow\uparrow})(n_\mathbf{q}+1)+ ....
\\
\hbar \frac{d}{dt}\rho^{\downarrow\uparrow}_{\downarrow\downarrow}=-\hbar \frac{d}{dt}\rho^{\uparrow\downarrow}_{\downarrow\downarrow}
&=&
\Gamma_d ( -\rho^{\downarrow\uparrow}_{\downarrow\downarrow}+\rho^{\uparrow\downarrow}_{\downarrow\downarrow})(n_\mathbf{q}+1)+ ....
\end{eqnarray}
\end{subequations}

Other terms in the Hamiltonian [Eq.~(\ref{eq:Hamiltonian_approx})] associated with $V_{l\mathbf{K}}$ are due to the interaction with contact electrons. Master equation derivations due to them do not involve boson particles and are essentially the same as those in, for example, Refs.~\cite{Dong_PRB04, Song_PRL14}. We do not repeat them here. The rest of terms, mainly due to $B$ field, are relatively easy to derive.

\section{Magnetic field modulated A-B chain}\label{sec:A-B chain}

For the cases that both impurities have energy level within the bias window, the quantum kinetic equations can be greatly simplified. There are four cases totally: A-A, B-B, A-B and B-A chains. In the following, let us focus on the A-B chain, that is, we set $\mu_L>\{E_{dL},E_{dR}+U_R\}>\mu_R$. Both $E_B$ and $k_B T$ are much smaller than the bias window.   To understand the system more clearly and in several steps, we first give results without magnetic field.
\begin{subequations}\label{eq:master_zeroB}
\begin{eqnarray}
\hbar\frac{d }{d t}\rho_{0\sigma}
&=&
-2\Gamma_L \rho_{0\sigma} + \Gamma_R \rho_{02},
\\
\hbar\frac{d }{d t}\rho_{02}
&=&
-2(\Gamma_L +\Gamma_R)\rho_{02}+2 \Gamma_d[-2n_\mathbf{q} \rho_{02}+ (n_\mathbf{q}+1)(\rho_{\uparrow\downarrow}+\rho_{\downarrow\uparrow}-2{\rm Re}\rho_{\uparrow\downarrow}^{\downarrow\uparrow})] ,
\\
\hbar\frac{d }{d t}\rho_{\sigma\bar\sigma}
&=&
\Gamma_L \rho_{0\bar\sigma} +\Gamma_R \rho_{\sigma2}
+2 \Gamma_d[n_\mathbf{q} \rho_{02}+ (n_\mathbf{q}+1)(-\rho_{\sigma\bar\sigma}+{\rm Re}\rho_{\sigma\bar\sigma}^{\bar\sigma \sigma})],
\\
\hbar\frac{d }{d t}\rho_{\sigma2}
&=&
 \Gamma_L \rho_{02} -2\Gamma_R \rho_{\sigma2},
 \\
\hbar\frac{d }{d t}\rho_{\sigma\sigma}
&=&
\Gamma_L \rho_{0\sigma} + \Gamma_R \rho_{\sigma2},
\\
\hbar\frac{d }{d t}\rho^{\downarrow\uparrow}_{\uparrow\downarrow}
&=&
\Gamma_d[-2n_\mathbf{q} \rho_{02}+ (n_\mathbf{q}+1)(\rho_{\uparrow\downarrow}+\rho_{\downarrow\uparrow}-2\rho_{\uparrow\downarrow}^{\downarrow\uparrow})]
\end{eqnarray}
and they are not all independent, but supplemented by
\begin{eqnarray}
1=\rho_{02}+\sum_\sigma (\rho_{0\sigma}+ \rho_{\sigma\bar\sigma} +\rho_{\sigma2} +\rho_{\sigma\sigma}),
\end{eqnarray}
\end{subequations}
where
$\Gamma_{\ell} =2\pi \sum_\mathbf{k} |V_{\ell \mathbf{k}}|^2 \delta(E_{d\ell}-\varepsilon_{\alpha\mathbf{k}})$ and $\Gamma_d = 2\pi \sum_\mathbf{q}|V_{dd}\lambda_q|^2\delta(|E_{dL}-E_{dR}|-\varepsilon_{\mathbf{q}})$, $n_\mathbf{q}=1/(\exp(\varepsilon_\mathbf{q}/k_B T)-1)$. Above we have assumed $E_{dL}>E_{dR}$. For $E_{dL}<E_{dR}$, just switch $n_\mathbf{q}$ and $n_\mathbf{q}+1$. Note that we have used the Fermi distribution of the electrons at the left(right) contact to be 1(0) in the bias window since $k_B T\ll eV$ in our case. The modification to include $f_{L,R}(\varepsilon)$ is straightforward.

We remark that the off-diagonal terms associated with $\Gamma_d$ are protect by the invariance of the system Hamiltonian in the spin space, and lead to coherence between the two impurities regardless of the approximations one has chosen. It is important for the equation system to be invariant upon rotation of spin coordination, otherwise it is not physical. The existence of these terms can be easily checked by applying unitary transformation to rotate the spin coordinate and requiring the resulting equation system unchanged.    From Eqs.~(\ref{eq:master_zeroB}) we can easily deduce that $\rho_{\uparrow\downarrow}=\rho_{\downarrow\uparrow}= \rho_{\uparrow\downarrow}^{\downarrow\uparrow}$ and all other elements are zero except $\rho_{\uparrow\uparrow}$ and $\rho_{\downarrow\downarrow}$. From the requirement that the system is invariant for rotation in spin space, we need to have
\begin{eqnarray}
\frac{1}{2}\rho_{\uparrow\uparrow}= \frac{1}{2}\rho_{\downarrow\downarrow}=\rho_{\uparrow\downarrow}=\rho_{\downarrow\uparrow}= \rho_{\uparrow\downarrow}^{\downarrow\uparrow}=\rho^{\uparrow\downarrow}_{\downarrow\uparrow}=\frac{1}{6},
\end{eqnarray}
which physically is the isotopic distribution in triplet state space [comprising one electron in each impurity; $\frac{1}{3}(J_z=-1)+\frac{1}{3}(J_z=0)+\frac{1}{3}(J_z=1)$]. Apparently, the current is blocked in this steady state. So the current-block state is imposed by the system symmetry and should therefore be robust.

Next we add two arbitrary $B$ fields at two impurity sites. Without loss of generality, we set $\theta_L=0$ and $\theta_R=\theta_R-\theta_L=\theta$. Again we assume $E_{dL}>E_{dR}$.
\begin{subequations}\label{eq:master_B}
\begin{eqnarray}
\hbar\frac{d }{d t}\rho_{0\sigma}
&=&
-2\Gamma_L \rho_{0\sigma} + \Gamma_R \rho_{02} -2E_{BR}\sin\theta {\rm Im}\rho^{0\bar\sigma}_{0\sigma},
\\
\hbar\frac{d }{d t}\rho_{02}
&=&
-2(\Gamma_L +\Gamma_R)\rho_{02}+2 \Gamma_d[-2n_\mathbf{q} \rho_{02}+ (n_\mathbf{q}+1)(\rho_{\uparrow\downarrow}+\rho_{\downarrow\uparrow}-2{\rm Re}\rho_{\uparrow\downarrow}^{\downarrow\uparrow})] ,
\\
\hbar\frac{d }{d t}\rho_{\sigma\bar\sigma}
&=&
\Gamma_L \rho_{0\bar\sigma} +\Gamma_R \rho_{\sigma2}
+2 \Gamma_d[n_\mathbf{q} \rho_{02}+ (n_\mathbf{q}+1)(-\rho_{\sigma\bar\sigma}+{\rm Re}\rho_{\sigma\bar\sigma}^{\bar\sigma \sigma})]
 +2E_{BR}\sin\theta {\rm Im}\rho^{\sigma\bar\sigma}_{\sigma\sigma},\label{eq:rate_rho_sigmabarsigma}
\\
\hbar\frac{d }{d t}\rho_{\sigma2}
&=&
 \Gamma_L \rho_{02} -2\Gamma_R \rho_{\sigma2},
 \\
\hbar\frac{d }{d t}\rho_{\sigma\sigma}
&=&
\Gamma_L \rho_{0\sigma} + \Gamma_R \rho_{\sigma2}
- E_{BR}\sin\theta {\rm Im}\rho^{\sigma\bar\sigma}_{\sigma\sigma},
\\
\hbar\frac{d }{d t}\rho^{\downarrow\uparrow}_{\uparrow\downarrow}
&=&
\Gamma_d[-2n_\mathbf{q} \rho_{02}+ (n_\mathbf{q}+1)(\rho_{\uparrow\downarrow}+\rho_{\downarrow\uparrow}-2\rho_{\uparrow\downarrow}^{\downarrow\uparrow})]
+
2i (-E_{BL}+E_{BR}\cos\theta)\rho^{\downarrow\uparrow}_{\uparrow\downarrow}
+i E_{BR}\sin\theta(\rho^{\downarrow\downarrow}_{\uparrow\downarrow}-\rho^{\downarrow\uparrow}_{\uparrow\uparrow})
\\
\hbar\frac{d }{d t}\rho^{0\downarrow}_{0\uparrow}
&=&
 -2\Gamma_L \rho^{0\downarrow}_{0\uparrow} + i E_{BR}[-2\cos\theta \rho^{0\downarrow}_{0\uparrow} +\sin\theta (\rho_{0\uparrow}-\rho_{0\downarrow}) ],
\\
\hbar\frac{d }{d t}\rho^{\sigma\bar\sigma}_{\sigma\sigma}
&=&
\Gamma_L \rho^{0\bar\sigma}_{0\sigma}+ \Gamma_d (n_\mathbf{q}+1)(\rho^{\bar\sigma\sigma}_{\sigma\sigma}-\rho^{\sigma\bar\sigma}_{\sigma\sigma})
  + i E_{BR}[-2\sigma\cos\theta \rho^{\sigma\bar\sigma}_{\sigma\sigma} +\sin\theta (\rho_{\sigma\sigma}-\rho_{\sigma\bar\sigma}) ],
\\
\hbar\frac{d }{d t}\rho^{\bar\sigma\sigma}_{\sigma\sigma}
&=&
\Gamma_L \rho^{\bar\sigma2}_{\sigma2}+ \Gamma_d (n_\mathbf{q}+1)(\rho^{\sigma\bar\sigma}_{\sigma\sigma}-\rho^{\bar\sigma\sigma}_{\sigma\sigma})
   -2\sigma i E_{BL}\rho^{\bar\sigma\sigma}_{\sigma\sigma} +i E_{BR}\sin\theta (\rho^{\bar\sigma\bar\sigma}_{\sigma\sigma}-\rho^{\bar\sigma\sigma}_{\sigma\bar\sigma}) ],
\\
\hbar\frac{d }{d t}\rho^{\downarrow2}_{\uparrow2}
&=&
-2\Gamma_R \rho^{\downarrow2}_{\uparrow2} -2 i E_{BL}\rho^{\downarrow2}_{\uparrow2},
\\
\hbar\frac{d }{d t}\rho^{\downarrow\downarrow}_{\uparrow\uparrow}
&=&
-2i (E_{BL}+E_{BR}\cos\theta) \rho^{\downarrow\downarrow}_{\uparrow\uparrow}
+i E_{BR} \sin\theta (\rho^{\downarrow\uparrow}_{\uparrow\uparrow}-\rho^{\downarrow\downarrow}_{\uparrow\downarrow}),
\end{eqnarray}
and they are not all independent and supplemented by
\begin{eqnarray}
1=\rho_{02}+\sum_\sigma (\rho_{0\sigma}+ \rho_{\sigma\bar\sigma} +\rho_{\sigma2} +\rho_{\sigma\sigma}),
\end{eqnarray}
\end{subequations}
We have solved the above system of equations and found the explicit analytical expression for all above density matrix elements. Since
\begin{eqnarray}\label{eq:current}
I = \frac{e}{\hbar}
2 \Gamma_L (\rho_{0\uparrow}+\rho_{0\downarrow}+\rho_{02})
=\frac{e}{\hbar}
2 \Gamma_R (\rho_{\uparrow2}+\rho_{\downarrow2}+\rho_{02}),
\end{eqnarray}
for the purpose of getting the current, we first show the the solutions for the following relevant elements
\begin{eqnarray}\label{eq:rho_02}
\rho_{02} &=& \frac{2\Gamma_L}{\Gamma_R}\rho_{0\uparrow}=\frac{2\Gamma_L}{\Gamma_R}\rho_{0\downarrow}
=\frac{2\Gamma_R}{\Gamma_L}\rho_{\uparrow2}=\frac{2\Gamma_R}{\Gamma_L}\rho_{\downarrow2}
\nonumber\\
&=&
2E_{BL}^2E_{BR}^2(E_{BL}^2-E_{BR}^2)^2 \Gamma_d\Gamma_L\Gamma_R  \sin^2\theta\bigg/
Deno
\nonumber\\
&=&
\frac{\Gamma_d\Gamma_L\Gamma_R  \sin^2\theta}
{ \Gamma_d(\Gamma_L^2+\Gamma_L\Gamma_R+\Gamma^2_R)\sin^2\theta +2\Gamma_L\Gamma_R(\Gamma_L+\Gamma_R)}, \; \textrm{if}\; \{E_{BL}, E_{BR}, |E_{BL}-E_{BR}|\}\gg\{\Gamma_R,\Gamma_L,\Gamma_d\},
\end{eqnarray}
 where we assume $n_q\ll 1$, and
\begin{eqnarray}\label{eq:deno}
Deno&=&
(E_{BL}^2+E_{BR}^2+2E_{BL}E_{BR}\cos\theta)\bigg[E_{BL}^4-E_{BL}^2E_{BR}^2(1+\cos^2\theta)+E_{BR}^4\bigg]
\Gamma^2_d\Gamma_L\Gamma_R(\Gamma_L+\Gamma_R)
\nonumber\\
&&
+2E_{BL}^2E_{BR}^2(E_{BL}^2-E_{BR}^2)^2 \bigg[\Gamma_d(\Gamma_L^2+\Gamma_L\Gamma_R+\Gamma^2_R)\sin^2\theta +2\Gamma_L\Gamma_R(\Gamma_L+\Gamma_R)\bigg].
\end{eqnarray}
Note that expression~(\ref{eq:rho_02}) is valid except at exactly $E_{BL}= E_{BR}$ which is a removable singularity point and $\rho_{02}=0$.

Realistically, impurities have effective internal magnetic field in addition to the external field \cite{Song_PRL14}, and $B$ field at each impurity is the vector sum of internal and external fields. Now we arrive at the results given in Ref.~\cite{Txoperena_PRL14} [see Eqs.(S2) and (S3) in the supplemental material in \cite{Txoperena_PRL14}]. Over there in that paper, we further integrated the internal field distribution and obtained the final dependence on the external field via a A-B impurity chain. Although it should keep the same trend under the applied magnetic field, we can further average over the locations and energies of the impurities. As a separate note, under the conditions $\{E_{BL}, E_{BR}, |E_{BL}-E_{BR}|\}\gg\{\Gamma_R,\Gamma_L,\Gamma_d\}$, the current from Eq.~(\ref{eq:current}) becomes that of a quasi-series resistor,
\begin{eqnarray}\label{eq:current_2}
I
\approx
\frac{2e}{\hbar} \left(
\frac{1}{\Gamma_L }+ \frac{1}{\Gamma_R }
-\frac{1}{ \Gamma_L+\Gamma_R}
 +\frac{4}{\Gamma_{d} \sin^2\theta}\right)^{-1},
\end{eqnarray}
as Eq.~(1) in Ref.~\cite{Txoperena_PRL14}. The third term $-\frac{1}{ \Gamma_L+\Gamma_R}$ is a typical correction to non-interacting hopping due to the fact that two electrons cannot always go through an impurity simultaneously in the large $U$ limit (a simpler situation also happens for Coulomb-repulsion-modified resonant tunneling as in Ref.~\cite{Glazman_resonant_JETP88}).

For other density matrix elements we have
\begin{subequations}\label{eq:rho_other}
\begin{eqnarray}
\rho_{\uparrow\uparrow} &=&  \rho_{\downarrow\downarrow}
\nonumber\\
&=&
\bigg\{
4E_{BL}^2E_{BR}^2(E_{BL}^2-E_{BR}^2)^2 (1+ \cos^2\theta)
+\bigg[2E_{BL}^6-3E_{BL}^4E_{BR}^2+2E_{BL}^2E_{BR}^4\sin^4\theta+E_{BR}^6
\nonumber\\
&&
\quad+E_{BR}(4 E_{BL}+E_{BR}\cos\theta)(E_{BL}^2-E_{BR}^2)^2\cos\theta
\bigg]\Gamma^2_d
\bigg\}
\Gamma_L\Gamma_R(\Gamma_L+\Gamma_R)
\bigg/
(4Deno)
\nonumber
\\
&=&\left\{
\begin{array}{cl}
\dfrac{ (1+ \cos^2\theta)\Gamma_L\Gamma_R(\Gamma_L+\Gamma_R)}
{2\Gamma_d(\Gamma_L^2+\Gamma_L\Gamma_R+\Gamma^2_R)\sin^2\theta
+4\Gamma_L\Gamma_R(\Gamma_L+\Gamma_R)}, \; &\textrm{if}\; \{E_{BL}, E_{BR}, |E_{BL}-E_{BR}|\}\gg\{\Gamma_R,\Gamma_L,\Gamma_d\},
\\
\frac{1}{2}\sin^2\frac{\theta}{2}, \; &\textrm{if}\; E_{BL}=E_{BR}, \theta\neq0
\\
\frac{1}{2}, \; &\textrm{if}\; E_{BL}\neq E_{BR}, \theta=0
\end{array}\right.\qquad
\end{eqnarray}

\begin{eqnarray}
\rho_{\uparrow\downarrow} &=&  \rho_{\downarrow\uparrow}
\nonumber\\
&=&
\sin^2\theta\bigg\{
E_{BR}^2\bigg[3E_{BL}^4+4E_{BL}^3E_{BR}\cos\theta-2E_{BL}^2E_{BR}^2\sin^2\theta+E_{BR}^4\bigg]
\Gamma^2_d
\nonumber\\
&&
\qquad\quad+
4E_{BL}^2E_{BR}^2(E_{BL}^2-E_{BR}^2)^2
\bigg\}
\Gamma_L\Gamma_R(\Gamma_L+\Gamma_R)
\bigg/
(4Deno)
\nonumber
\\
&=&\left\{
\begin{array}{cl}
\dfrac{ \sin^2\theta
\Gamma_L\Gamma_R(\Gamma_L+\Gamma_R)}
{2\Gamma_d(\Gamma_L^2+\Gamma_L\Gamma_R+\Gamma^2_R)\sin^2\theta
+4\Gamma_L\Gamma_R(\Gamma_L+\Gamma_R)}, \; &\textrm{if}\; \{E_{BL}, E_{BR}, |E_{BL}-E_{BR}|\}\gg\{\Gamma_R,\Gamma_L,\Gamma_d\},
\\
\frac{1}{2}\cos^2\frac{\theta}{2}, \; &\textrm{if}\; E_{BL}=E_{BR},\theta\neq0
\\
0, \; &\textrm{if}\; E_{BL}\neq E_{BR}, \theta=0
\end{array}\right.\qquad
\end{eqnarray}
\begin{eqnarray}
\rho^{\downarrow\uparrow}_{\uparrow\downarrow}\!\!
&=&\!\!
\left[(3E_{BL}^4+4E_{BL}^3E_{BR}\cos\theta\!-\!2E_{BL}^2E_{BR}^2\sin^2\theta\!+\!E_{BR}^4)
\Gamma_d
-2i
E_{BL} (E_{BL}^2\!-\!E_{BR}^2) (2E_{BL}^2+E_{BL}E_{BR}\cos\theta\!-\!E_{BR}^2)\right]
\nonumber\\
&&
\sin^2\theta E_{BR}^2\Gamma_d\Gamma_L\Gamma_R(\Gamma_L+\Gamma_R)
\bigg/(4Deno)
\nonumber
\\
&=&\left\{
\begin{array}{cl}
0, \; &\textrm{if}\; \{E_{BL}, E_{BR}, |E_{BL}-E_{BR}|\}\gg\{\Gamma_R,\Gamma_L,\Gamma_d\},
\\
\frac{1}{2}\cos^2\frac{\theta}{2}, \; &\textrm{if}\; E_{BL}=E_{BR},\theta\neq0
\\
0, \; &\textrm{if}\; E_{BL}\neq E_{BR}, \theta=0
\end{array}\right.
\end{eqnarray}
\begin{eqnarray}
\rho^{\downarrow\downarrow}_{\uparrow\uparrow}
&=&
\left[(E_{BL}^4-2E_{BL}^2E_{BR}^2(1+\sin^2\theta)+E_{BR}^4)
\Gamma_d
-2i
E_{BL} E_{BR}(E_{BL}^2-E_{BR}^2) (E_{BL}\cos\theta-E_{BR})\right] \qquad\qquad\qquad\qquad
\nonumber\\
&&
\sin^2\theta E_{BR}^2 \Gamma_d\Gamma_L\Gamma_R(\Gamma_L+\Gamma_R)
\bigg/(4Deno)
\nonumber
\\
&=&\left\{
\begin{array}{cl}
0, \; &\textrm{if}\; \{E_{BL}, E_{BR}, |E_{BL}-E_{BR}|\}\gg\{\Gamma_R,\Gamma_L,\Gamma_d\},
\\
-\frac{1}{2}\sin^2\frac{\theta}{2}, \; &\textrm{if}\; E_{BL}=E_{BR},
\end{array}\right.
\end{eqnarray}
\begin{eqnarray}
\rho^{\uparrow\downarrow}_{\uparrow\uparrow}&=& -\rho^{\downarrow\downarrow}_{\downarrow\uparrow}
\nonumber\\
&=&
\bigg\{ 4E_{BL}^2E_{BR}^2 (E_{BL}^2-E_{BR}^2)^2 \cos\theta
+E_{BR}(E_{BL}+E_{BR}\cos\theta)\left[(E_{BL}^2-E_{BR}^2)^2-2E_{BL}^2E_{BR}^2\sin^2\theta \right]
\Gamma^2_d \qquad\qquad
\nonumber\\
&&
\quad+2i
E^2_{BL} E_{BR}(E_{BL}^2-E_{BR}^2)^2\Gamma_d \bigg\}
\sin\theta\Gamma_L\Gamma_R(\Gamma_L+\Gamma_R)
\bigg/(4Deno)
\nonumber
\\
&=&\left\{
\begin{array}{cl}
\dfrac{\cos\theta
\sin\theta\Gamma_L\Gamma_R(\Gamma_L+\Gamma_R)}
{2\Gamma_d(\Gamma_L^2+\Gamma_L\Gamma_R+\Gamma^2_R)\sin^2\theta
+4\Gamma_L\Gamma_R(\Gamma_L+\Gamma_R)}, \; &\textrm{if}\; \{E_{BL}, E_{BR}, |E_{BL}-E_{BR}|\}\gg\{\Gamma_R,\Gamma_L,\Gamma_d\},
\\
-\frac{1}{2}\sin\frac{\theta}{2}\cos\frac{\theta}{2}, \; &\textrm{if}\; E_{BL}=E_{BR},
\end{array}\right.
\end{eqnarray}
\begin{eqnarray}
\rho^{\downarrow\uparrow}_{\uparrow\uparrow}&=& -\rho^{\downarrow\downarrow}_{\uparrow\downarrow}
\nonumber\\
&=&
\bigg\{E_{BR}(E_{BL}+E_{BR}\cos\theta)\left[(E_{BL}^2-E_{BR}^2)^2-2E_{BL}^2E_{BR}^2\sin^2\theta \right]
\Gamma^2_d
\nonumber\\
&&
\quad-2i
E_{BL} E^2_{BR}(E_{BL}^2-E_{BR}^2)[(E_{BL}^2-E_{BR}^2)\cos\theta-E_{BL}E_{BR}\sin^2\theta]\Gamma_d \bigg\}
\sin\theta\Gamma_L\Gamma_R(\Gamma_L+\Gamma_R)
\bigg/(4Deno)
\nonumber
\\
&=&\left\{
\begin{array}{cl}
0, \; &\textrm{if}\; \{E_{BL}, E_{BR}, |E_{BL}-E_{BR}|\}\gg\{\Gamma_R,\Gamma_L,\Gamma_d\},
\\
-\frac{1}{2}\sin\frac{\theta}{2}\cos\frac{\theta}{2},  \; &\textrm{if}\; E_{BL}=E_{BR},
\end{array}\right.
\\
\rho^{0\downarrow}_{0\uparrow}&=&\rho^{\downarrow2}_{\uparrow2}
=0
\end{eqnarray}
\end{subequations}

With density matrix results in Eqs.~(\ref{eq:rho_02}) and (\ref{eq:rho_other}), we now analyze the current-block solution  at $E_{BL}=E_{BR}$ (but $\theta\neq 0$). We can transform it ($\rho_{B_L}$, spin quantization along $\mathbf{B}_L$) into the spin basis that along the bisector of two $B$ field directions, $\rho_{bis}$.
\begin{eqnarray}
\rho_{bis} = U_{\frac{\theta}{2}} \rho_{B_L} U^T_{\frac{\theta}{2}},
\end{eqnarray}
where from above
\begin{eqnarray}
\rho_{B_L}=\left(\renewcommand{\arraystretch}{1.5} \begin{array}{cccc}
\frac{1}{2}\sin^2\frac{\theta}{2} & -\frac{1}{2}\sin\frac{\theta}{2}\cos\frac{\theta}{2} & -\frac{1}{2}\sin\frac{\theta}{2}\cos\frac{\theta}{2} &-\frac{1}{2}\sin^2\frac{\theta}{2}
\\
 -\frac{1}{2}\sin\frac{\theta}{2}\cos\frac{\theta}{2} & \frac{1}{2}\cos^2\frac{\theta}{2} & \frac{1}{2}\cos^2\frac{\theta}{2} & \frac{1}{2}\sin\frac{\theta}{2}\cos\frac{\theta}{2}
\\
 -\frac{1}{2}\sin\frac{\theta}{2}\cos\frac{\theta}{2} & \frac{1}{2}\cos^2\frac{\theta}{2} & \frac{1}{2}\cos^2\frac{\theta}{2} & \frac{1}{2}\sin\frac{\theta}{2}\cos\frac{\theta}{2}
 \\
-\frac{1}{2}\sin^2\frac{\theta}{2} & \frac{1}{2}\sin\frac{\theta}{2}\cos\frac{\theta}{2} & \frac{1}{2}\sin\frac{\theta}{2}\cos\frac{\theta}{2} &\frac{1}{2}\sin^2\frac{\theta}{2}
\end{array}
\right),
\end{eqnarray}
has non-zero elements within the subspace of $\{|\uparrow_{B_L}\uparrow_{B_L}\rangle, |\uparrow_{B_L}\downarrow_{B_L}\rangle, |\downarrow_{B_L}\uparrow_{B_L}\rangle,   |\downarrow_{B_L}\downarrow_{B_L}\rangle\}$, and $U_{\frac{\theta}{2}}$ is the unitary matrix that rotates a state vector by $\frac{\theta}{2}$ (from spin quantization along $\mathbf{B}_L$ to that along the bisector)
\begin{eqnarray}
U_{\frac{\theta}{2}}=\left(\renewcommand{\arraystretch}{1.5} \begin{array}{cccc}
\cos^2\frac{\theta}{4} & \sin\frac{\theta}{4}\cos\frac{\theta}{4} & \sin\frac{\theta}{4}\cos\frac{\theta}{4} &\sin^2\frac{\theta}{4}
\\
 -\sin\frac{\theta}{4}\cos\frac{\theta}{4} & \cos^2\frac{\theta}{4} & -\sin^2\frac{\theta}{4} & \sin\frac{\theta}{4}\cos\frac{\theta}{4}
\\
 -\sin\frac{\theta}{4}\cos\frac{\theta}{4} & -\sin^2\frac{\theta}{4} & \cos^2\frac{\theta}{4} & \sin\frac{\theta}{4}\cos\frac{\theta}{4}
 \\
\sin^2\frac{\theta}{4} & -\sin\frac{\theta}{4}\cos\frac{\theta}{4} & -\sin\frac{\theta}{4}\cos\frac{\theta}{4} &\cos^2\frac{\theta}{4}
\end{array}
\right).
\end{eqnarray}
Then, we have
\begin{eqnarray}\label{eq:rho_bis}
\rho_{bis}=\left(\renewcommand{\arraystretch}{1.2} \begin{array}{cccc}
0&0&0&0
\\
 0&\frac{1}{2} & \frac{1}{2}&0
\\
 0&\frac{1}{2} & \frac{1}{2}&0
 \\
0&0&0&0
\end{array}
\right),
\end{eqnarray}
with the basis of $\{|\uparrow_{bis}\uparrow_{bis}\rangle, |\uparrow_{bis}\downarrow_{bis}\rangle, |\downarrow_{bis}\uparrow_{bis}\rangle,   |\downarrow_{bis}\downarrow_{bis}\rangle\}$. It is physically a $\{\mathbf{J}=1, J_z=0\}$ state. Note that this result holds regardless of the value of $n_q$ and has been explicitly checked.  This state is a symmetrically allowed state for this particular $E_{BL}=E_{BR}$ configuration. Apparently, the current is blocked at this solution. This result is somewhat a surprise, as the $B$ fields at  the two impurities generally point in different directions. As it satisfies the system symmetry in spin space, this result is directly related to the coherent off-diagonal terms in the master equations. This term has an observable  effect: as one increases the applied magnetic field from 0, the hopping current  increase a little before decreasing. This is what the calculated curve shows in Ref.~\cite{Txoperena_PRL14} [Fig. 3c by integrating Eq.(S6) of that supplemental material and considering $I=0$ at $E_{BL}=E{BR}$]. This behavior was seen in some of the experimental samples in Ref.~\cite{Txoperena_PRL14}, but we have not unambiguously identified it with the coherent effect. A specifically designed experiment may better probe this effect.

Another current-block solution is at $\theta=0$ (but $E_{BL}\neq E_{BR}$), and
\begin{eqnarray}\label{eq:rho_B}
\rho_{B}=\left(\renewcommand{\arraystretch}{1.2} \begin{array}{cccc}
\frac{1}{2}&0&0&0
\\
 0&0 & 0&0
\\
 0&0 &0&0
 \\
0&0&0&\frac{1}{2}
\end{array}
\right).
\end{eqnarray}
This is a mixed state that consists equally $\{\mathbf{J}=1, J_z=1\}$ and $\{\mathbf{J}=1, J_z=-1\}$ two components, which is symmetry allowed in spin space. This state represents the limit of infinitely large external field and results in the current-blocked solution. Therefore, this current blockade of the B-A chain is again imposed by the system symmetry and is therefore very robust.   The reason that $\rho_{B,\uparrow\uparrow}= \rho_{B,\downarrow\downarrow}$ (this is generally true regardless of $\theta$) rather than $\rho_{B,\uparrow\uparrow}< \rho_{B,\downarrow\downarrow}$ is that in our case $g\mu B\ll k_B T$ and thus the thermal equilibration between the Zeeman split states is negligible. The spin-flip mechanisms like hyperfine interaction on the impurities have effectively been taken into account by the internal field. Even $\rho_{B,\uparrow\uparrow}< \rho_{B,\downarrow\downarrow}$ if we are in $g\mu B \geq k_B T$ regime and take into account the thermal relaxation, this symmetry-allowed state remains current blockade. In all cases,  the current-block  state we reach is physically intuitive: once the system randomly selects this state it cannot evolve out of itself, and therefore the steady state will be finally locked into this (symmetry-allowed) current-block state.

The solution at exactly $\theta=0$ \& $E_{BL}= E_{BR}$ is not well defined. The specific physical insight  for that the triplet state reduces to $J_z=0$ [Eq.~(\ref{eq:rho_bis})] as $E_{BL}= E_{BR}$ and $J_z=\pm 1$ [Eq.~(\ref{eq:rho_B})] as $\theta=0$ is an open question to us.

\section{Magnetic field modulated B-A, A-A, and B-B chains}\label{sec:other chains}

The following derivation leads to part of the results in the supplemental material of Ref.~\cite{Txoperena_PRL14}.
\subsection{B-A chain}\label{subsec:B-A chains}

Repeat the above procedure for the B-A chain case, we get unblocked current independent of $B$ fields. Without loss of generality, we set $\theta_L=0$ and $\theta_R=\theta_R-\theta_L=\theta$. Again we assume $E_{dL}>E_{dR}$.
\begin{subequations}\label{eq:master_BA}
\begin{eqnarray}
\hbar\frac{d }{d t}\rho_{\sigma0}
&=&
-\Gamma_L \rho_{\sigma0} + \Gamma_R (\rho_{\sigma\sigma}+\rho_{\sigma\bar{\sigma}}),
\\
\hbar\frac{d }{d t}\rho_{20}
&=&
\Gamma_L(\rho_{\uparrow0}+\rho_{\downarrow0}) + \Gamma_R(\rho_{2\uparrow}+\rho_{2\downarrow})
+2 \Gamma_d[n_\mathbf{q} (\rho_{\uparrow\downarrow}+\rho_{\downarrow\uparrow}-2{\rm Re}\rho_{\uparrow\downarrow}^{\downarrow\uparrow})- 2\rho_{20}(n_\mathbf{q}+1)] ,
\\
\hbar\frac{d }{d t}\rho_{\sigma\bar\sigma}
&=&
-(\Gamma_L  +\Gamma_R) \rho_{\sigma\bar\sigma}
+ 2 \Gamma_d[ n_\mathbf{q}(-\rho_{\sigma\bar\sigma}+{\rm Re}\rho_{\sigma\bar\sigma}^{\bar\sigma \sigma}) + (n_\mathbf{q}+1)\rho_{20}]
 +2E_{BR}\sin\theta {\rm Im}\rho^{\sigma\bar\sigma}_{\sigma\sigma},
\\
\hbar\frac{d }{d t}\rho_{2\sigma}
&=&
 \Gamma_L(\rho_{\sigma\sigma}+\rho_{\bar\sigma \sigma})  -\Gamma_R \rho_{2\sigma}-2 E_{BR} \sin\theta {\rm Im}\rho^{2\bar\sigma}_{2\sigma},
 \\
\hbar\frac{d }{d t}\rho_{\sigma\sigma}
&=&
-(\Gamma_L  + \Gamma_R) \rho_{\sigma\sigma}
- E_{BR}\sin\theta {\rm Im}\rho^{\sigma\bar\sigma}_{\sigma\sigma},
\\
\hbar\frac{d }{d t}\rho^{\downarrow\uparrow}_{\uparrow\downarrow}
&=&
-(\Gamma_L  + \Gamma_R)\rho^{\downarrow\uparrow}_{\uparrow\downarrow}
+\Gamma_d[n_\mathbf{q}(\rho_{\uparrow\downarrow}+\rho_{\downarrow\uparrow}-2\rho_{\uparrow\downarrow}^{\downarrow\uparrow})
-2(n_\mathbf{q}+1) \rho_{20}]
\nonumber\\
&&+
2i (-E_{BL}+E_{BR}\cos\theta)\rho^{\downarrow\uparrow}_{\uparrow\downarrow}
+i E_{BR}\sin\theta(\rho^{\downarrow\downarrow}_{\uparrow\downarrow} - \rho^{\downarrow\uparrow}_{\uparrow\uparrow})
\\
\hbar\frac{d }{d t}\rho^{2\downarrow}_{2\uparrow}
&=&
 \Gamma_L (\rho^{\uparrow\downarrow}_{\uparrow\uparrow} + \rho^{\downarrow\downarrow}_{\downarrow\uparrow})+ \Gamma_R \rho^{2\downarrow}_{2\uparrow} + i E_{BR}[-2\cos\theta \rho^{2\downarrow}_{2\uparrow} +\sin\theta (\rho_{2\uparrow}-\rho_{2\downarrow}) ],
\\
\hbar\frac{d }{d t}\rho^{\sigma\bar\sigma}_{\sigma\sigma}
&=&
-(\Gamma_L +\Gamma_R)\rho^{\sigma\bar\sigma}_{\sigma\sigma}
+ \Gamma_d n_\mathbf{q}(\rho^{\bar\sigma\sigma}_{\sigma\sigma}-\rho^{\sigma\bar\sigma}_{\sigma\sigma})
  + i E_{BR}[-2\sigma\cos\theta \rho^{\sigma\bar\sigma}_{\sigma\sigma} +\sin\theta (\rho_{\sigma\sigma}-\rho_{\sigma\bar\sigma}) ],
\\
\hbar\frac{d }{d t}\rho^{\bar\sigma\sigma}_{\sigma\sigma}
&=&
-(\Gamma_L +\Gamma_R)\rho^{\bar\sigma\sigma}_{\sigma\sigma}
+ \Gamma_d  n_\mathbf{q} (\rho^{\sigma\bar\sigma}_{\sigma\sigma}-\rho^{\bar\sigma\sigma}_{\sigma\sigma})
   -2\sigma i E_{BL}\rho^{\bar\sigma\sigma}_{\sigma\sigma} +i E_{BR}\sin\theta (\rho^{\bar\sigma\bar\sigma}_{\sigma\sigma}-\rho^{\bar\sigma\sigma}_{\sigma\bar\sigma}) ],
\\
\hbar\frac{d }{d t}\rho^{\downarrow\downarrow}_{\uparrow\uparrow}
&=&
-(\Gamma_L +\Gamma_R)\rho^{\downarrow\downarrow}_{\uparrow\uparrow}
-2i (E_{BL}+E_{BR}\cos\theta) \rho^{\downarrow\downarrow}_{\uparrow\uparrow}
+i E_{BR} \sin\theta (\rho^{\downarrow\uparrow}_{\uparrow\uparrow}-\rho^{\downarrow\downarrow}_{\uparrow\downarrow}),
\end{eqnarray}
and they are not all independent and supplemented by
\begin{eqnarray}
1=\rho_{20}+\sum_\sigma (\rho_{\sigma0}+ \rho_{\sigma\bar\sigma} +\rho_{2\sigma} +\rho_{\sigma\sigma}),
\end{eqnarray}
\end{subequations}

The current thus is
\begin{eqnarray}
I = \frac{e}{\hbar}
\Gamma_L \sum_\sigma(\rho_{\sigma0}+\rho_{\sigma\bar\sigma}+\rho_{\sigma\sigma})
=\frac{e}{\hbar}
 \Gamma_R \sum_\sigma(\rho_{2\sigma}+\rho_{\sigma\bar\sigma}+\rho_{\sigma\sigma}).
\end{eqnarray}
and the diagonal matrix elements are
\begin{subequations}
\begin{eqnarray}
\rho_{\sigma0}  &=&
\frac{2 \Gamma_d\Gamma_R^2}{\Lambda},
\quad
\rho_{2\sigma}  =
\frac{2 \Gamma_d\Gamma_L^2}{\Lambda},
\quad
\rho_{20}  =
\frac{\Gamma_L\Gamma_R(\Gamma_L+\Gamma_R)}{\Lambda},
\\
\rho_{\sigma\sigma}  &=&
\frac{4E_{BR}^2\sin^2\theta}{4E_{BR}^2+(\Gamma_L+\Gamma_R)^2}
\frac{\Gamma_d\Gamma_L\Gamma_R}{\Lambda},
\\
\rho_{\sigma\bar\sigma}  &=&
\left[1-\frac{2E_{BR}^2\sin^2\theta}{4E_{BR}^2+(\Gamma_L+\Gamma_R)^2}\right]
\frac{2\Gamma_d\Gamma_L\Gamma_R}{\Lambda},
\end{eqnarray}
\end{subequations}
where we assume $n_{\mathbf{q}}\ll 1$, and
\begin{eqnarray}
\Lambda&=&
\Gamma_L\Gamma_R(\Gamma_L+\Gamma_R)+4\Gamma_d(\Gamma_L^2+\Gamma_L\Gamma_R+\Gamma^2_R).
\end{eqnarray}
So, the current is field-independent,
\begin{eqnarray}
I = \frac{e}{\hbar}
\frac{4\Gamma_d\Gamma_L\Gamma_R(\Gamma_L+\Gamma_R) }{\Gamma_L\Gamma_R(\Gamma_L+\Gamma_R)+4\Gamma_d(\Gamma_L^2+\Gamma_L\Gamma_R+\Gamma^2_R)}
.
\end{eqnarray}
In fact, by setting $E_{BL}=E_{BR}=0$ in Eq.~(\ref{eq:master_BA}), we can easily get the field-independent current expression for arbitrary $n_\mathbf{q}$,
\begin{eqnarray}
I = \frac{e}{\hbar}
\frac{4\Gamma_d\Gamma_L\Gamma_R(\Gamma_L+\Gamma_R) (n_\mathbf{q}+1) }{\Gamma_L\Gamma_R(\Gamma_L+\Gamma_R+4\Gamma_d n_\mathbf{q})+4\Gamma_d(\Gamma_L^2+\Gamma_L\Gamma_R+\Gamma^2_R)(n_\mathbf{q}+1)}
.
\end{eqnarray}
For completeness, the off-diagonal elements solution of Eq.~(\ref{eq:master_BA}) are
\begin{subequations}
\begin{eqnarray}
\rho^{\downarrow\uparrow}_{\uparrow\uparrow}  &=& -\rho^{\downarrow\downarrow}_{\uparrow\downarrow}
=
\sin^2\theta E^2_{BR}(\Gamma_L+\Gamma_R)
\frac{4\Gamma_d\Gamma_L\Gamma_R}{\Lambda\times deno2 }
\nonumber\\
&&
\qquad\quad\{
(\Gamma_L+\Gamma_R)[12E^2_{BL}-4E^2_{BR}-(\Gamma_L+\Gamma_R)^2]+2iE_{BL} [-4E^2_{BL}+4E^2_{BR}+3(\Gamma_L+\Gamma_R)^2]\},
\\
\rho^{\uparrow\downarrow}_{\uparrow\uparrow}  &=& -\rho^{\downarrow\uparrow}_{\downarrow\downarrow}
=-\frac{\sin\theta E_{BR}(2E_{BR}\cos\theta+i)}{4E_{BR}^2+(\Gamma_L+\Gamma_R)^2}
\frac{2\Gamma_d\Gamma_L\Gamma_R}{\Lambda},
\\
\rho^{\downarrow\uparrow}_{\uparrow\uparrow}  &=& -\rho^{\downarrow\downarrow}_{\uparrow\downarrow}
=
\sin\theta E_{BR}(\Gamma_L+\Gamma_R)
\frac{2\Gamma_d\Gamma_L\Gamma_R}{\Lambda\times deno2}
\{
 [4E_{BL}^2+(\Gamma_L+\Gamma_R)^2]\Lambda_3
+2E_{BR}\cos\theta  (\Gamma_L+\Gamma_R) \Lambda_4 \},
\\
\rho^{\downarrow\uparrow}_{\uparrow\downarrow}  &=&
(\Gamma_L+\Gamma_R)
\frac{2\Gamma_d\Gamma_L\Gamma_R}{\Lambda\times deno2}
 \bigg( 2E^2_{BR}\sin^2\theta   \Lambda_4+ [4E_{BL}^2+(\Gamma_L+\Gamma_R)^2]\times
\nonumber\\
&&
\!\!\!\!\!\!\!\!\!\!\!\!
\big\{-(\Gamma_L+\Gamma_R)[4(E_{BL}^2+E_{BR}^2)+(\Gamma_L+\Gamma_R)^2]+2i E_{BL}[4(E_{BL}^2-E_{BR}^2)+(\Gamma_L+\Gamma_R)^2]
-2 E_{BR}\cos\theta\Lambda_3\big\}
\bigg)
\end{eqnarray}
where
\begin{eqnarray}
deno2&=& [4E_{BR}^2+(\Gamma_L+\Gamma_R)^2]
[16(E_{BL}^2-E_{BR}^2)^2+8(E_{BL}^2+E_{BR}^2)(\Gamma_L+\Gamma_R)^2+(\Gamma_L+\Gamma_R)^4],
\\
\Lambda_3&=&4(\Gamma_L+\Gamma_R)E_{BL}-i[4 (E_{BL}^2 - E^2_{BR})-(\Gamma_L+\Gamma_R)^2],
\\
\Lambda_4&=& (\Gamma_L+\Gamma_R)[12E_{BL}^2-4E^2_{BR}-(\Gamma_L+\Gamma_R)^2] +2i  E_{BL} [-4E_{BL}^2+4E^2_{BR}+3(\Gamma_L+\Gamma_R)^2]
\end{eqnarray}
\end{subequations}

\subsection{A-A chain}\label{subsec:A-A chains}

For the A-A chain case, we get unblocked current, as well as density matrix elements, independent of $B$ fields. Without loss of generality, we set $\theta_L=0$ and $\theta_R=\theta_R-\theta_L=\theta$. Again we assume $E_{dL}>E_{dR}$.
\begin{subequations}\label{eq:master_AA}
\begin{eqnarray}
\hbar\frac{d }{d t}\rho_{\sigma0}
&=&
\Gamma_L \rho_{00} + \Gamma_R (\rho_{\sigma\sigma}+\rho_{\sigma\bar{\sigma}}) +2\Gamma_d[n_\mathbf{q}\rho_{0\sigma}-(n_\mathbf{q}+1)\rho_{\sigma0}],
\\
\hbar\frac{d }{d t}\rho_{0\sigma}
&=&
-(2\Gamma_L+\Gamma_R) \rho_{0\sigma} +2\Gamma_d[-n_\mathbf{q}\rho_{0\sigma}+(n_\mathbf{q}+1)\rho_{\sigma0}]
 -2E_{BR}\sin\theta {\rm Im}\rho^{0\bar\sigma}_{0\sigma},
\\
\hbar\frac{d }{d t}\rho_{00}
&=&
-2\Gamma_L \rho_{00} + \Gamma_R(\rho_{0\uparrow}+\rho_{0\downarrow}),
\\
\hbar\frac{d }{d t}\rho_{\sigma\sigma}
&=&
\Gamma_L \rho_{0\sigma}  - \Gamma_R\rho_{\sigma\sigma}
-2 E_{BR}\sin\theta {\rm Im}\rho^{\sigma\bar\sigma}_{\sigma\sigma},
\\
\hbar\frac{d }{d t}\rho_{\sigma\bar\sigma}
&=&
\Gamma_L \rho_{0\bar\sigma}-\Gamma_R \rho_{\sigma\bar\sigma}
 +2E_{BR}\sin\theta {\rm Im}\rho^{\sigma\bar\sigma}_{\sigma\sigma},
\\
\hbar\frac{d }{d t}\rho^{0\downarrow}_{0\uparrow}
&=&
-(2 \Gamma_L +\Gamma_R)\rho^{0\downarrow}_{0\uparrow} +2\Gamma_d[-n_\mathbf{q}\rho^{0\downarrow}_{0\uparrow}+(n_\mathbf{q}+1)\rho^{\downarrow0}_{\uparrow0}]+ i E_{BR}[-2\cos\theta \rho^{0\downarrow}_{0\uparrow} +\sin\theta (\rho_{0\uparrow}-\rho_{0\downarrow}) ],
\\
\hbar\frac{d }{d t}\rho^{\downarrow0}_{\uparrow0}
&=&
\Gamma_R(\rho^{\downarrow\uparrow}_{\uparrow\uparrow} +\rho^{\downarrow\downarrow}_{\uparrow\downarrow}) +2\Gamma_d[n_\mathbf{q}\rho^{0\downarrow}_{0\uparrow}-(n_\mathbf{q}+1)\rho^{\downarrow0}_{\uparrow0}]-2 i E_{BL}\rho^{\downarrow0}_{\uparrow0},
\\
\hbar\frac{d }{d t}\rho^{\sigma\bar\sigma}_{\sigma\sigma}
&=&
\Gamma_L\rho^{0\bar\sigma}_{0\sigma} -\Gamma_R \rho^{\sigma\bar\sigma}_{\sigma\sigma}
  + i E_{BR}[-2\sigma\cos\theta \rho^{\sigma\bar\sigma}_{\sigma\sigma} +\sin\theta (\rho_{\sigma\sigma}-\rho_{\sigma\bar\sigma}) ],
\\
\hbar\frac{d }{d t}\rho^{\bar\sigma\sigma}_{\sigma\sigma}
&=&
-\Gamma_R\rho^{\bar\sigma\sigma}_{\sigma\sigma}
   -2\sigma i E_{BL}\rho^{\bar\sigma\sigma}_{\sigma\sigma} +i E_{BR}\sin\theta (\rho^{\bar\sigma\bar\sigma}_{\sigma\sigma}-\rho^{\bar\sigma\sigma}_{\sigma\bar\sigma}) ],
\\
\hbar\frac{d }{d t}\rho^{\downarrow\downarrow}_{\uparrow\uparrow}
&=&
-\Gamma_R\rho^{\downarrow\downarrow}_{\uparrow\uparrow}
-2i (E_{BL}+E_{BR}\cos\theta) \rho^{\downarrow\downarrow}_{\uparrow\uparrow}
+i E_{BR} \sin\theta (\rho^{\downarrow\uparrow}_{\uparrow\uparrow}-\rho^{\downarrow\downarrow}_{\uparrow\downarrow}),
\\
\hbar\frac{d }{d t}\rho^{\downarrow\uparrow}_{\uparrow\downarrow}
&=&
- \Gamma_R\rho^{\downarrow\uparrow}_{\uparrow\downarrow}
+2i (-E_{BL}+E_{BR}\cos\theta)\rho^{\downarrow\uparrow}_{\uparrow\downarrow}
+i E_{BR}\sin\theta(\rho^{\downarrow\downarrow}_{\uparrow\downarrow} - \rho^{\downarrow\uparrow}_{\uparrow\uparrow}),
\end{eqnarray}
and they are not all independent and supplemented by
\begin{eqnarray}
1=\rho_{00}+\sum_\sigma (\rho_{\sigma0} +\rho_{0\sigma} +\rho_{\sigma\sigma}+ \rho_{\sigma\bar\sigma}),
\end{eqnarray}
\end{subequations}
The solutions are
\begin{eqnarray}
\rho_{0\sigma} &=&\frac{\Gamma_R}{\Gamma_L}\rho_{\sigma'\sigma''}= \frac{\Gamma_L}{\Gamma_R}\rho_{00} = \frac{2\Gamma_d(n_\mathbf{q}+1)}{2\Gamma_L+\Gamma_R+2\Gamma_d n_\mathbf{q}} \rho_{\sigma'0}
\nonumber\\
&=&
\frac{\Gamma_L\Gamma_R\Gamma_d (n_\mathbf{q}+1)}{ (4\Gamma_L^2+2\Gamma_L\Gamma_R+\Gamma_R^2)\Gamma_d(n_\mathbf{q}+1) +\Gamma_L\Gamma_R(2\Gamma_L+ \Gamma_R +2\Gamma_d n_\mathbf{q})},
\end{eqnarray}
and all the off-diagonal elements are 0.
The current is
\begin{eqnarray}
I &=& \frac{e}{\hbar}
2\Gamma_L (\rho_{00}+\sum_\sigma \rho_{0\sigma}
=\frac{e}{\hbar}
 \Gamma_R \sum_\sigma(\rho_{0\sigma}+\rho_{\sigma\bar\sigma}+\rho_{\sigma\sigma})
 \nonumber\\
&=&2\frac{e}{\hbar}(\Gamma_R+2\Gamma_L)\rho_{0\uparrow}
 \nonumber\\
&=& \frac{e}{\hbar}\frac{2 (\Gamma_R+2\Gamma_L) \Gamma_L\Gamma_R \Gamma_d(n_\mathbf{q}+1)}
{(4\Gamma_L^2+2\Gamma_L\Gamma_R+\Gamma_R^2)\Gamma_d(n_\mathbf{q}+1) +\Gamma_L\Gamma_R(2\Gamma_L+\Gamma_R + 2\Gamma_d n_\mathbf{q})}
 \nonumber\\
&\approx&
\frac{e}{\hbar}\frac{2 (\Gamma_R+2\Gamma_L) \Gamma_L\Gamma_R \Gamma_d}
{(4\Gamma_L^2+2\Gamma_L\Gamma_R+\Gamma_R^2)\Gamma_d +\Gamma_L\Gamma_R(2\Gamma_L+\Gamma_R) },\quad{\rm if} \; n_\mathbf{q}\ll 0.
\end{eqnarray}

\subsection{B-B chain}\label{subsec:B-B chains}

For the B-B chain case, we get unblocked current, as well as density matrix elements, independent of $B$ fields. Without loss of generality, we set $\theta_L=0$ and $\theta_R=\theta_R-\theta_L=\theta$. Again we assume $E_{dL}>E_{dR}$.

\begin{subequations}\label{eq:master_AA}
\begin{eqnarray}
\hbar\frac{d }{d t}\rho_{2\sigma}
&=&
\Gamma_L(\rho_{\sigma\sigma}+\rho_{\bar{\sigma}\sigma})   + \Gamma_R\rho_{22} +2\Gamma_d[n_\mathbf{q}\rho_{\sigma2}-(n_\mathbf{q}+1)\rho_{2\sigma}]
-2E_{BR}\sin\theta {\rm Im}\rho^{2\bar\sigma}_{2\sigma},
\\
\hbar\frac{d }{d t}\rho_{\sigma2}
&=&
-(\Gamma_L+2\Gamma_R) \rho_{\sigma2} +2\Gamma_d[-n_\mathbf{q}\rho_{\sigma2}+(n_\mathbf{q}+1)\rho_{2\sigma}],
\\
\hbar\frac{d }{d t}\rho_{22}
&=&
 \Gamma_L(\rho_{\uparrow2}+\rho_{\downarrow2})-2\Gamma_R \rho_{22},
\\
\hbar\frac{d }{d t}\rho_{\sigma\sigma}
&=&
-\Gamma_L \rho_{\sigma\sigma} + \Gamma_R\rho_{\sigma2}
- 2E_{BR}\sin\theta {\rm Im}\rho^{\sigma\bar\sigma}_{\sigma\sigma},
\\
\hbar\frac{d }{d t}\rho_{\sigma\bar\sigma}
&=&
-\Gamma_L\rho_{\sigma\bar\sigma} +\Gamma_R \rho_{\sigma2}
 +2E_{BR}\sin\theta {\rm Im}\rho^{\sigma\bar\sigma}_{\sigma\sigma},
\\
\hbar\frac{d }{d t}\rho^{\downarrow2}_{\uparrow2}
&=&
-( \Gamma_L +2\Gamma_R)\rho^{\downarrow2}_{\uparrow2} +2\Gamma_d[-n_\mathbf{q}\rho^{\downarrow2}_{\uparrow2}+(n_\mathbf{q}+1)\rho^{2\downarrow}_{2\uparrow}]
-2 i E_{BL}\rho^{\downarrow2}_{\uparrow2},
\\
\hbar\frac{d }{d t}\rho^{2\downarrow}_{2\uparrow}
&=&
\Gamma_L(\rho^{\uparrow\downarrow}_{\uparrow\uparrow} +\rho^{\downarrow\downarrow}_{\downarrow\uparrow}) +2\Gamma_d[n_\mathbf{q}\rho^{\downarrow2}_{\uparrow2}-(n_\mathbf{q}+1)\rho^{2\downarrow}_{2\uparrow}]
+ i E_{BR}[-2\cos\theta \rho^{2\downarrow}_{2\uparrow} +\sin\theta (\rho_{2\uparrow}-\rho_{2\downarrow}) ]
\\
\hbar\frac{d }{d t}\rho^{\sigma\bar\sigma}_{\sigma\sigma}
&=&
-\Gamma_L \rho^{\sigma\bar\sigma}_{\sigma\sigma}
  + i E_{BR}[-2\sigma\cos\theta \rho^{\sigma\bar\sigma}_{\sigma\sigma} +\sin\theta (\rho_{\sigma\sigma}-\rho_{\sigma\bar\sigma}) ],
\\
\hbar\frac{d }{d t}\rho^{\bar\sigma\sigma}_{\sigma\sigma}
&=&
-\Gamma_L\rho^{\bar\sigma\sigma}_{\sigma\sigma}+ \Gamma_R\rho^{\bar\sigma2}_{\sigma2}
   -2\sigma i E_{BL}\rho^{\bar\sigma\sigma}_{\sigma\sigma} +i E_{BR}\sin\theta (\rho^{\bar\sigma\bar\sigma}_{\sigma\sigma}-\rho^{\bar\sigma\sigma}_{\sigma\bar\sigma}) ],
\\
\hbar\frac{d }{d t}\rho^{\downarrow\downarrow}_{\uparrow\uparrow}
&=&
-\Gamma_L\rho^{\downarrow\downarrow}_{\uparrow\uparrow}
-2i (E_{BL}+E_{BR}\cos\theta) \rho^{\downarrow\downarrow}_{\uparrow\uparrow}
+i E_{BR} \sin\theta (\rho^{\downarrow\uparrow}_{\uparrow\uparrow}-\rho^{\downarrow\downarrow}_{\uparrow\downarrow}),
\\
\hbar\frac{d }{d t}\rho^{\downarrow\uparrow}_{\uparrow\downarrow}
&=&
- \Gamma_L\rho^{\downarrow\uparrow}_{\uparrow\downarrow}
+2i (-E_{BL}+E_{BR}\cos\theta)\rho^{\downarrow\uparrow}_{\uparrow\downarrow}
+i E_{BR}\sin\theta(\rho^{\downarrow\downarrow}_{\uparrow\downarrow} - \rho^{\downarrow\uparrow}_{\uparrow\uparrow}),
\end{eqnarray}
and they are not all independent and supplemented by
\begin{eqnarray}
1=\rho_{22}+\sum_\sigma (\rho_{\sigma2} +\rho_{2\sigma} +\rho_{\sigma\sigma}+ \rho_{\sigma\bar\sigma}),
\end{eqnarray}
\end{subequations}
The solutions are
\begin{eqnarray}
\rho_{\sigma2} &=&\frac{\Gamma_L}{\Gamma_R}\rho_{\sigma'\sigma''}= \frac{\Gamma_R}{\Gamma_L}\rho_{22} = \frac{2\Gamma_d(n_\mathbf{q}+1)}{\Gamma_L+2\Gamma_R+2\Gamma_d n_\mathbf{q}} \rho_{2\sigma'}
\nonumber\\
&=&
\frac{\Gamma_L\Gamma_R\Gamma_d (n_\mathbf{q}+1)}
{ (\Gamma_L^2+2\Gamma_L\Gamma_R+4\Gamma_R^2)\Gamma_d(n_\mathbf{q}+1) +\Gamma_L\Gamma_R(\Gamma_L+ 2\Gamma_R +2\Gamma_d n_\mathbf{q})},
\end{eqnarray}
and all the off-diagonal elements are 0.
The current is
\begin{eqnarray}
I &=& \frac{e}{\hbar}
2\Gamma_R (\rho_{22}+\sum_\sigma \rho_{\sigma2}
=\frac{e}{\hbar}
 \Gamma_L \sum_\sigma(\rho_{\sigma2}+\rho_{\sigma\bar\sigma}+\rho_{\sigma\sigma})
 \nonumber\\
&=&2\frac{e}{\hbar}(2\Gamma_R+\Gamma_L)\rho_{\uparrow2}
 \nonumber\\
&=& \frac{e}{\hbar}\frac{2 (2\Gamma_R+\Gamma_L) \Gamma_L\Gamma_R \Gamma_d(n_\mathbf{q}+1)}
{(\Gamma_L^2+2\Gamma_L\Gamma_R+4\Gamma_R^2)\Gamma_d(n_\mathbf{q}+1) +\Gamma_L\Gamma_R(\Gamma_L+2\Gamma_R + 2\Gamma_d n_\mathbf{q})}
 \nonumber\\
&\approx&
\frac{e}{\hbar}\frac{2 (2\Gamma_R+\Gamma_L) \Gamma_L\Gamma_R \Gamma_d}
{(\Gamma_L^2+2\Gamma_L\Gamma_R+4\Gamma_R^2)\Gamma_d +\Gamma_L\Gamma_R(\Gamma_L+2\Gamma_R )},\quad{\rm if} \; n_\mathbf{q}\ll 0.
\end{eqnarray}

\section{Summary}\label{sec:summary}

We have shown the rigorous derivations for the magnetic-field-modulated tunnel current by phonon-assisted hopping through a A-B impurity chain, as well as the $B$-independent current via a B-A, A-A or A-B chain. The full analytical solutions of the master equations are presented. The obtained coherent off-diagonal terms in the master equations are discussed with the symmetry of the system Hamiltonian in spin space. It may lead to an observable effect at small applied magnetic field where the tunnel current via a A-B chain slightly increases before decreasing at larger $B$ field.  Our TMR effect works in the regime of $\{g\mu B, k_B T \}\ll eV$. It distinguishes itself from other TMR mechanisms that often require strong magnetic field, small bias window and/or low temperature, such as the tunnel junction version of the Kondo effect \cite{Appelbaum_PRL66, Goldhaber_Nat98}, or the Coulomb-interaction-modified resonant tunneling or hopping \cite{Glazman_resonant_JETP88, Ephron_PRL92, Bahlouli_PRB94}.

\section*{Acknowledgements}
I would like to thank Professor Hanan Dery for insightful discussions and F\`{e}lix Casanova for the related experimental results.


\begin{thebibliography}{99}

\bibitem{Txoperena_PRL14} O. Txoperena, Y. Song, L. Qing, M. Gobbi, L. E. Hueso, H. Dery, and F\`{e}lix Casanova, arXiv:1404.0633.
\bibitem{Song_PRL14} Y. Song and H. Dery, Phys. Rev. Lett. \textbf{113},047205 (2014).
\bibitem{Mahato_Science13} R. N. Mahato, H. L\"{u}lf, M. H. Siekman, S. P. Kersten, P. A. Bobbert, M. P. de Jong, L. De Cola, and W. G. van der Wiel, Science \textbf{341}, 257 (2013).

\bibitem{Glazman_phonon_JETP88} L. I. Glazman and K. A. Matveev, , Zh. Eksp. Teor. Fiz. \textbf{94}, 332 (1988) [Sov. Phys. JETP \textbf{67}, 1276 (1988)].
\bibitem{Xu_PRB95} Y. Xu, D. Ephron, and M. R. Beasley, Phys. Rev. B \textbf{52},2843 (1995).

\bibitem{Bahlouli_PRB94} H. Bahouli, K. A. Matveev, D. Ephron, and M. R. Beasley, Phys. Rev. B \textbf{49}, 14496 (1994).


\bibitem{Abrikosov_Phy65} A. A. Abrikosov, Physics (Long Island City, N.Y.) \textbf{2}, 21 (1965).
\bibitem{Barnes_JPhys76} S. E. Barnes, J. Phys. F: et. Phys. \textbf{6}, 1375 (1976).
\bibitem{Coleman_PRB84} P. Coleman, Phys. Rev. B \textbf{29}, 3035 (1984).
\bibitem{Zou_PRB88} Z. Zou and P. W. Anderson, Phys. Rev. B \textbf{37}, 627(R) (1988).
\bibitem{LeGuillou_PRB95} J. C. Le Guillou and E. Ragoucy, Phys. Rev. B \textbf{52}, 2403 (1995).

\bibitem{Langreth_chap76} D. C. Langreth, \textit{in Linear and Nonlinear Electron Transport in Solids}, Vol. \textbf{17} \textit{of Nato Advanced Study Institute, Series B: Physics}, edited by J. T. Devreese and V. E. Van Doren (Plenum, New York, 1976).
\bibitem{Haug_Jauho_book08} H. J. W. Haug and A.-P. Jauho, \textit{Quantum Kinetics in Transport and Optics of Semiconductors} (Springer, 2008).

\bibitem{Dong_PRB04} B. Dong, H. L. Cui, and X. L. Lei, Phys. Rev. B \textbf{69}, 035324 (2004).

\bibitem{Appelbaum_PRL66} J. Appelbaum, Phys. Rev. Lett. \textbf{17}, 91 (1966).
\bibitem{Goldhaber_Nat98}D. Goldhaber-Gordon, H. Shtrikman, D. Mahalu, D. Abusch-Magder, U. Meirav and M. A. Kastner, Nature \textbf{391}, 156 (1998).
\bibitem{Glazman_resonant_JETP88} L. I. Glazman and K. A. Matveev, Pis'ma Zh. Eksp. Teor. Fiz. \textbf{48}, 403 (1988) [JETP Lett. \textbf{48}, 445 (1988)].
\bibitem{Ephron_PRL92} D. Ephron, Y. Xu, and M. R. Beasley, Phys. Rev. Lett. \textbf{3112}, 69 (1992).
\end{thebibliography}
\end{document}